\documentclass[aip,graphicx]{revtex4-1}
\usepackage{graphicx}
\usepackage{dcolumn}
\usepackage{bm}
\usepackage{color} %

\draft 

\begin{document}


\title{Rich structural phase diagram and thermoelectric properties of layered tellurides Mo$_{1-x}$Nb$_{x}$Te$_{2}$} 



\author{Koji Ikeura}
\affiliation{Department of Applied Physics, University of Tokyo, Hongo, Tokyo 113-8656, Japan}

\author{Hideaki Sakai}
\affiliation{Department of Applied Physics, University of Tokyo, Hongo, Tokyo 113-8656, Japan}

\author{Mohammad Saeed Bahramy}
\affiliation{Department of Applied Physics, University of Tokyo, Hongo, Tokyo 113-8656, Japan}
\affiliation{RIKEN Center for Emergent Matter Science (CEMS), Wako, Saitama 351-0198, Japan}

\author{Shintaro Ishiwata}
\email[]{ishiwata@ap.t.u-tokyo.ac.jp}
\affiliation{Department of Applied Physics, University of Tokyo, Hongo, Tokyo 113-8656, Japan}
\affiliation{JST, PRESTO, Kawaguchi, Saitama 332-0012, Japan}

\date{\today}

\begin{abstract}
MoTe$_2$ is a rare transition-metal ditelluride having two kinds of layered polytypes, hexagonal structure with trigonal prismatic Mo coordination and monoclinic structure with octahedral Mo coordination. The monoclinic distortion in the latter is caused by anisotropic metal-metal bonding. In this work, we have examined the Nb doping effect on both polytypes of MoTe$_2$ and clarified a structural phase diagram for Mo$_{1-x}$Nb$_{x}$Te$_{2}$ containing four kinds of polytypes. A rhombohedral polytype crystallizing in polar space group has been newly identified as a high-temperature metastable phase at slightly Nb-rich composition.  Considering the results of thermoelectric measurements and the first principles calculations, the Nb ion seemingly acts as a hole dopant in the rigid band scheme. On the other hand, the significant interlayer contraction upon the Nb doping, associated with the Te $p$-$p$ hybridization, is confirmed especially for the monoclinic phase, which implies a shift of the $p$-band energy level. The origin of the metal-metal bonding in the monoclinic structure is discussed in terms of the $d$ electron counting and the Te $p$-$p$ hybridization. 
\end{abstract}


\maketitle 

Transition metal dichalcogenides (TMDs) $MX_2$ ($X$ = S, Se, Te) with layered structure have been extensively studied due to their wide variety of electronic properties such as superconductivity, large thermoelectric effect, and anomalous magnetoresistance.\cite{Wilson, TiS2, WTe2} In addition, recent experimental progress on TMDs exfoliated down to a monolayer or a few layers have stimulated great interest in their potential applications to spin-valleytronics and optoelectronics.\cite{Wang,Xu} When containing heavy elements and lacking inversion symmetry, a strong spin-orbit coupling yields the interplay of spin and valley degrees of freedom. 

The diverse electronic properties of layered TMDs are enriched by structural polymorphism, which are differentiated by the chalcogen coordination of metal ion and by the stacking sequence of $MX_2$ layers.\cite{Wilson, Jobic1992, Pod} The polytypes commonly obtained are 1T (CdI$_2$-type structure) and 2H structures, consisting of the edge-shared octahedra and the trigonal prisms of $MX_6$, respectively (see Fig. 1). The integer of the polytypes denotes the number of $MX_2$ layers per unit cell and the capital letter stands for the symmetry of Bravais lattice such as trigonal (T), hexagonal (H), and rhombohedral (R). Spurred by the seminal works on monolayer MoS$_2$,\cite{Mak_monolayer,Xiao_monolayer} 3R-MoS$_2$ without inversion symmetry has been revealed to show the valley-dependent photoluminescence even in multilayers.\cite{Suzuki} 

The group V and VI transition-metal (TM) ditellurides crystallize mostly in 1T polytype. Among them, MoTe$_2$ is an exceptional TM ditelluride adopting two kinds of polytypes, 2H polytype (so called $\alpha$-MoTe$_2$) \cite{Puotinen} and distorted 1T polytype (so called $\beta$-MoTe$_2$).\cite{Brown} $\alpha$-MoTe$_2$ shows semiconducting behavior and transforms into the high-temperature metallic phase $\beta$-MoTe$_2$ at 820$-$880 $^{\circ}$C.\cite{Revo1966,Vell1970} In $\alpha$-MoTe$_2$, the trigonal prismatic coordination causes the crystal field splitting for a non-bonding $d$ band, by which a narrow band formed by $d_{z^2}$ orbitals splits off. Given that the $p$-$d$ hybridization is negligibly small, the nominal valence of the molybdenum ion is 4+ with a $d^2$ electron counting. Thus, the Fermi level is expected to reside inside the gap with fully occupying the $d_{z^2}$ orbital.\cite{Dawson,Boker} On the other hand, the structural and electronic properties of $\beta$-MoTe$_2$, which can be obtained as a metastable phase by quenching from above 900 $^{\circ}$C, are not so straightforward. $\beta$-MoTe$_2$ at room temperature crystallizes in monoclinically distorted 1T polytype ($P$2$_1$/$m$) with semimetallic band structure.\cite{Brown} In each MoTe$_2$ layer, zigzag chains made of metal-metal bonding are formed along the $b$ axis, giving rise to a considerable shift of Mo ions along the $a$ axis from the center of each octahedron (See Fig. 3(a)). The metal-metal bonding is also observed in NbTe$_{2}$ having distorted 1T structure ($C$2/$m$) with double zigzag chains.\cite{Brown_NbTe2} In this paper, we renamed the distorted 1T polytypes for $\beta$-MoTe$_2$ and NbTe$_2$ as 1T' and 1T" polytypes, respectively (note that the number of 1T'-MoTe$_2$ layers per unit cell is 2). The origin of the chain formations has been discussed in terms of the Fermi surface nesting in the $d$ band and electron-phonon coupling.\cite{Whangbo,NbTe2} However, the situation is not so simple because of a strong $p$-$d$ hybridization inherent to tellurides with 1T structure.

In this work, structure and transport properties of Mo$_{1-x}$Nb$_{x}$Te$_{2}$ are studied to see how Nb doping modifies lattice and electronic structures of the two polytypes, 1T'- and 2H-MoTe$_{2}$. We have established a detailed structural phase diagram containing the newly identified 3R polytype. Considering the thermoelectric measurements and first principles calculations, the effect of the Nb doping is simply interpreted in terms of the rigid band shift of the Fermi level. On the other hand, an enhancement of the Te $p$-$p$ hybridization affecting the valence band structure is signified by a significant interlayer contraction especially for Nb-doped 1T'-MoTe$_{2}$ with metal-metal bonding. We discuss the important role of the Te $p$ orbitals as well as the $d$ electron counting in the structure and electronic properties of Nb-doped MoTe$_{2}$. 

All polycrystalline samples were synthesized by solid state reaction in evacuated quartz tubes. The mixtures of stoichiometric amounts of elements, Mo (purity 99.9 $\%$), Nb (purity 99.9 $\%$), and Te (purity 99.999 $\%$), in powder form were heated at 1050 - 1100 $^{\circ}$C for 12 h, followed by cooling down to room temperature in 3 h.  The obtained samples were ground, pelletized, and then annealed at selected temperatures between 800 and 1100 $^{\circ}$C for 12 h. The annealed samples were obtained after water quenching, except for 2H-MoTe$_2$ which was cooled to room temperature by furnace cooling. Powder x-ray diffraction (XRD) patterns were corrected on a Bruker D8 advance diffractometer using Cu $K$$\alpha$ radiation. For transport measurements, the samples were cut into rectangular shape in the typical dimensions of 5 $\times$ 1 $\times$ 0.3 mm$^3$. The thermopower and thermal conductivity were measured by a conventional steady-state method with a temperature difference of less than 1 K between the voltage electrodes. 

Electronic structure calculations were performed within the context of density functional theory using the Perdew-Burke-Ernzerhof exchange-correlation functional modified by the Becke-Johnson potential, as implemented in the WIEN2K program  \cite{Blaha}. Relativistic effects, including spin-orbit coupling, were fully included. The muffin-tin radius of each atom $R_{\rm{MT}}$ was chosen such that its product with the maximum modulus of reciprocal vectors $K_{\rm{max}}$ become $R_{\rm{MT}}K_{\rm{max}} =7.0$. The structural parameters were taken from the reference \cite{Dawson}. For 2H-MoTe$_{2}$  (1T'-MoTe$_{2}$), the corresponding Brillouin zone was sampled by a $15\times15\times 5$ ($8\times16\times 4$) $k$-mesh. 

Selected XRD patterns of Mo$_{1-x}$Nb$_{x}$Te$_{2}$ annealed at  800 $^{\circ}$C and 1050 $^{\circ}$C are shown in Figs. 1(a) and 1(b), respectively.  The XRD patterns indicate the successful syntheses of a whole solid solution (weak reflections of an unidentified impurity phase are seen for compounds with $x$ = 0.6$-$0.8). The solid solution contains four kinds of structures, 2H, 1T', 1T", and newly identified 3R polytype. Lattice parameters and space group for the selected compounds are summarized in Table I. 

Let us describe the important features of XRD profiles for samples annealed at 1050 $^{\circ}$C. The $h$0$l$ reflections around 2$\theta \sim$ 35$^{\circ}$ are useful to discriminate the 3R phase from the 2H phase. However, those reflections for $x$ = 0.4$-$0.6 suffer from a significant broadening, while the 00$l$ reflections remain sharp. It has been reported that Ta$_{1-x}$$M_x$Se$_2$ ($M$ = Re, Os), of which $d$ electron counting deviates from an integer number, has a mixed-layer structure consisting of an incoherent stacking of 2H and 3R layers. The volume fraction of 2H and 3R varies depending on the chemical composition.\cite{Hayashi} The mixed-layer structure (or equivalently intergrowth) is detectable as the peak broadening of the $h$0$l$ reflections originated from either 2H or 3R polytypes. The same feature is found for the XRD patterns of compounds with $x$ = 0.4$-$0.6 obtained by annealing at 1050 $^{\circ}$C. The existence of the mixed-layer structure in the present system reflects the proximity of the free energies between the 2H and 3R polytypes in addition to the entropy contribution at high temperatures. Besides the peak broadening except for the 00$l$ reflections, the similar peak broadening is also seen for the 1T' phase with $x$ = 0 and 0.2, indicating the mixed-layer structure with a small amount of 2H layer forming during the quenching process. 

Figure 2 shows a structural phase diagram of Mo$_{1-x}$Nb$_{x}$Te$_{2}$ as functions of $x$ and the annealing temperature. Despite the rare formation of the trigonal prismatic coordination in the tellurides, 2H and 3R polytypes appear in a wide composition range from $x$ = 0 up to $x$ = 0.7 with completely destabilizing the other phases in the range 0.3 $< x <$ 0.7. This feature is in contrast to the robust stability of the undistorted 1T phase in sulfides and selenides with a non-integer $d$ electron counting such as (Nb,Rh)(S,Se)$_2$, (Nb,Ir)S$_2$, and (Ti,Ir)Se$_2$.\cite{Maeda,Shimakawa_NbIrS,Shimakawa_TiIrSe}  The undistorted 1T structure  ($P\overline{3}m$1) potentially existing at high temperatures have not been obtained presumably because of the unquenchable nature or the incompatibility with the $d$ electron counting of $d^1-d^2$. The 3R polytype exists as a meta-stable state in a narrow temperature-composition range and is more stable than the 2H polytype at high temperatures. As mentioned above, the 3R phase obtained in this study contains a minor portion of the 2H layers as mixed-layers. The relative stability between 2H and 3R polytypes has been discussed for sulfides and selenides. The theoretical studies for MoS$_2$ and MoSe$_2$ have pointed out that the 2H and 3R phases are energetically almost degenerate but the former is more favorable than the latter.\cite{He} It has been reported for MoS$_2$ that the 2H phase transforms into the 3R phase at high pressures above 4 GPa and at high temperatures above 1900 $^{\circ}$C.\cite{Silverman} Accordingly, the extent of the stable composition for 3R-Mo$_{1-x}$Nb$_{x}$Te$_{2}$ is expected to be enlarged by the application of high pressure as well. This expectation is supported by the experimental fact that 3R-Mo$_{0.4}$Nb$_{0.6}$Te$_{2}$ has a higher crystal density compared to that of 2H polytype in the same composition.

To study the relation between the structural and the electronic properties of Mo$_{1-x}$Nb$_{x}$Te$_{2}$, structural parameters calculated from the lattice constants are plotted as a function of Nb content $x$ in Figs. 3(d)-(j). Note that the lattice parameters for compounds with a minor phase or the mixed-layer structure were refined as two phase system, and the data of the major phase only were presented in the plots. As confirmed in Figs. 3(i) and 3(j), the unit cell volume per formula unit ($V$/$Z$) of Mo$_{1-x}$Nb$_{x}$Te$_{2}$ increases almost linearly with increasing $x$, reflecting the larger ionic radius of the Nb ion compared with the Mo ion. 

As shown in Figs. 3(a)-(c), 1T' and 1T" layers have a modulated triangular lattice with inplane distortion, whereas 2H and 3R layers have an equilateral triangular (or hexagonal) lattice without the distortion. The inplane distortion in Fig. 3(d) is a dimensionless index, which is defined by $a/\sqrt{3}b$ for 1T' and $a/3\sqrt{3}b$ for 1T" structures, so that the corresponding one for 2H and 3R structures, $a'/\sqrt{3}a$, becomes unity. Since the inplane distortion is caused by the formation of the zigzag chains (Figs. 3(a) and 3(c)), the significant deviation of the inplane distortion from the unity suggests that the metal-metal bonding remains intact throughout 1T' and 1T" phases. 

Contrary to the volume expansion, considerable contraction of the interplane distance is observed especially for 1T' polytype. As $x$ increases from 0 to 0.3, the interplane distance decreases by 1.2 $\%$ for 2H polytype and 2.3 $\%$ for 1T' polytype (Figs. 3(g) and 3(h)), while their inplane lattice constants increase monotonically (Figs. 3(e) and 3(f)). This behavior can be accounted for by considering the interlayer $p$-$p$ hybridization and the $p$-$d$ charge transfer affecting the anisotropic bonding in this system, i.e., the weak interlayer interaction mediated by van der Waals force and the strong inplane interaction mediated by covalent and ionic bonding, as will be discussed later.

Figure 4 shows temperature dependence of thermoelectric properties for selected compounds covering all four polytypes. Resistivity and Seebeck coefficient of 1T'-MoTe$_{2}$ show a kink at 230 K, which corresponds to the first order structural transition from a high-temperature monoclinic phase ($P$2$_1$/$m$) to a low-temperature orthorhombic phase ($Pnm$2$_1$).\cite{Clarke,Hughes}  2H-MoTe$_{2}$ has a Seebeck coefficient of -385 $\mu$V/K (data not shown) and a resistivity of about 10$^4$ times larger value than that of the Nb 10$\%$-doped compound at room temperature. The negative Seebeck coefficient with semiconducting behavior (data not shown) implies the presence of a slight Te deficiency in 2H-MoTe$_{2}$. All Nb-doped compounds show metallic behavior and positive Seebeck coefficient in the entire temperature range (see Figs.4(a)-(d)). No superconductivity was observed down to 2 K, but 1T"-NbTe$_2$ shows a slight down turn in resistivity near the lowest measured temperature. It is notable that 1T'-Mo$_{0.9}$Nb$_{0.1}$T$_{2}$ shows an anomalous increase in Seebeck coefficient as temperature decreases with a maxima at $\sim$50 K.  This behavior is in marked contrast to diffusive thermopower in conventional metallic system. The hump structure in thermal conductivity below 100 K both for 2H and 1T' tends to be smeared out by the chemical substitution which suppresses the increment of the phonon mean free path at low temperatures (see Figs.4(e) and 4(f)).

The composition-dependent thermoelectric properties at 300 K are summarized in Fig. 5. As shown in Fig. 5(d), 1T'- and 2H-MoTe$_{2}$ are found to maximize their thermoelectric figures of merit $ZT \sim$0.02 at 300 K by Nb doping of about 10 $\%$, whereas $ZT$ for 1T" polytype is one order of magnitude lower. Both for semiconducting 2H-MoTe$_{2}$ and semimetallic 1T'-MoTe$_{2}$, resistivity decreases monotonically as indicated by the blue (2H) and red (1T') lines as guides to the eyes. Correspondingly, positive Seebeck coefficient for 2H polytype decreases systematically on going from $x$ = 0.1 to 0.6. These results suggest that the Nb ion acts as a hole dopant. As for 1T' polytype, the increase of the positive Seebeck coefficient by 10$\%$-Nb doping presumably reflects the disappearance of the electron pocket in 1T'-MoTe$_{2}$ with semimetallic band structure. To further ensure the interpretation of the experimental results, we have performed first-principles calculations as presented in Fig. 6 and estimated the Seebeck coefficient within the semiclassical Boltzmann theory. The calculations for Seebeck coefficient (averaged over all directions) were performed in the constant scattering time approximation where the lattice parameters are adopted for pristine MoTe$_2$. The effect of Nb doping was treated within the rigid band approximation by shifting down the chemical potential of pristine MoTe$_2$ to an appropriate energy corresponding to the given hole concentration. Seebeck coefficients at 300 K as a function of $x$ both for 2H- and 1T'-Mo$_{1-x}$Nb$_{x}$Te$_{2}$ are qualitatively consistent with the calculated values. The 10-$\%$ Nb doping corresponds to the shift of the Fermi level downward in the rigid-band approximation for 1T'(2H)-MoTe$_{2}$ by 125 meV (170 meV), where a peaky structure appears in the density of states (DOS) of the $p_x$/$p_y$ orbitals in 1T' polytype and that of the $p_z$ orbital in 2H polytype, leading to the enhanced Seebeck coefficient for 1T'- and 2H-Mo$_{0.9}$Nb$_{0.1}$Te$_{2}$. Consistent with the thermoelectric measurements, the band calculation has confirmed the disappearance of electron pockets by 10-$\%$ Nb doping for 1T'-MoTe$_{2}$ (data not shown). However, especially for 1T' polytype, the rigid-band approximation may be violated by the significant interlayer contraction as described below.

 Finally, we mention the effect of the Nb doping on the electronic and crystal structures of 2H- and 1T'-MoTe$_2$. It has been reported for TM ditellurides that the contraction of the van der Waals gap is associated with the increase in hybridization between Te $p_z$ orbitals (see the inset of Fig.3(g)), which raises the energy level of the antibonding Te $p_z$ band located at the top of the valence band.\cite{Te-Te} Likewise, in Fig. 3(h) the significant contraction of the interplane distance with increasing $x$ suggests the strengthening of the Te $p$-$p$ hybridization upon the Nb doping, giving rise to the enhancement of a three dimensional character. Here, we presume that the Mo-Te bonds in 2H polytype are rather ionic and the valence of Mo and Nb ions are 4+ and (4-$\delta$)+, respectively. Given this presumption, the contraction of the interplane distance upon the Nb doping in the 2H polytype can be interpreted as the charge redistribution from ($M^{4+}$, Te$^{2-}$) to ($M^{(4-x\delta)+}$, Te$^{(2-x\delta/2)-}$), where the enhanced $p$-$p$ hybridization compensates the charge deficiency of Te ions. On the other hand, for 1T' polytype, the effect of Nb doping on the interplane distance is even more significant, which implies an additional effect arising from the zigzag chain formation. The concept of the hidden Fermi surface nesting has been proposed as the electronic origin of the zigzag chain formation \cite{Whangbo}. Suppose that two out of three $t_{2g}$ orbitals of Mo$^{4+}$ ions accommodate one electron per each and the orbitals form two sets of independent one-dimensional chains within each layer. Then, the each chain would suffer from the Peierls instability intrinsic to the half-filled state, eventually resulting in the zigzag chain formation in the presence of the phonon instability and the electron-phonon coupling. Although the Nb substitution for Mo tends to destabilize the zigzag chain formation by reducing the $d$ electron counting, it is presumable that the strong $p$-$p$ hybridization between Te-Te double layers preserves the $d$ electron counting to prevent the chain from the destabilization to some extent through the $p$-$d$ charge transfer. Simply speaking, the Nb doping for the 1T' phase tends to create a hole in the Te $p_z$ band, so that the apparent charge distribution of Mo$_{1-x}$Nb$_{x}$Te$_{2}$ can be described as ($M^{(4-x)+}$, Te$^{(2-x/2)-}$). This mechanism accounts for the more pronounced contraction of the interplane distance upon the Nb doping in the 1T' phase, compared to the 2H phase, i.e., the averaged $d$ electron counting in the 1T' phase is relatively robust against the Nb doping. To further elucidate the primary origin of the zigzag chain formation, first-principles calculations for electron and phonon band structure of the Nb doped system is indispensable.



In summary, structural phase diagram of complete solid solution Mo$_{1-x}$Nb$_{x}$Te$_{2}$ has been established. The 3R polytype was identified for the first time as a TM ditelluride, as far as we know. Structural studies revealed that the Nb doping for Mo significantly increases the extent of the $p$-$p$ hybridization in between Te-Te double layers. This tendency is more pronounced in the 1T' phase compared to the 2H phase. Therefore, although the thermoelectric measurements, combined with the theoretical calculations, apparently indicate that the Nb ion substituted for Mo ion acts as a hole dopant in the rigid-band approximation, a sizable modification of the valence band structure should be considered to understand the effect of Nb doping especially for the 1T' phase, where the zigzag Mo chain requires a certain $d$ electron counting. Our work demonstrates that the structural chemistry and the electronic properties of TM ditellurides can be more fruitful by chemically tuning the $d$ electron counting or the interlayer $p$-$p$ hybridization.



%

\begin{acknowledgments}
The authors thank Y. Tokura, K. Ishizaka, and H. Tsuji for fruitful discussions. This study was partly supported by the Japan Society for the Promotion of Science (JSPS) Grants-in-Aid for Scientific Research (Nos. 25620040 and 24224009), the Thermal $\&$ Electric Energy Technology Inc. Foundation, and JST PRESTO program (Hyper-nano-space design toward Innovative Functionality).
\end{acknowledgments}


\begin{table}[h]
\caption{\label{TABLE I.} Structure parameters for four kinds of polytypes of Mo$_{1-x}$Nb$_{x}$Te$_{2}$ at room temperature. }
\begin{tabular}{cccccccc}
\hline
      & space group \ & \ $Z$ \ & \ $a$ (\AA)   & $b$ (\AA)      & $c$ (\AA)     & $\beta$ ($^{\circ}$)     &   $V$ (\AA$^3$)         \\ \hline
1T'-MoTe$_2$        & $P$2$_1$/$m$   & 4 & 6.3126(4)  & 3.4711(2) & 13.8052(8) & 93.726(5) & 301.86(3) \\ 
2H-MoTe$_2$         & $P$6$_3$/$mmc$ & 2 & 3.51943(5) &     -     & 13.9670(3) &     -      &   149.823(5)  \\ 
3R-Mo$_{0.4}$Nb$_{0.6}$Te$_2$ & $R$3$m$     & 3 & 3.6015(3)  & -         & 20.468(3)  & -   &   229.92(5)    \\ 
1T"-NbTe$_2$        & $C$2/$m$    & 6 & 19.321(1)  & 3.6336(2) & 9.3077(4)  & 134.589(2) & 465.36(4) \\ \hline
\end{tabular}
\end{table}

\newpage

 \begin{figure}
 \includegraphics[keepaspectratio,width=16 cm]{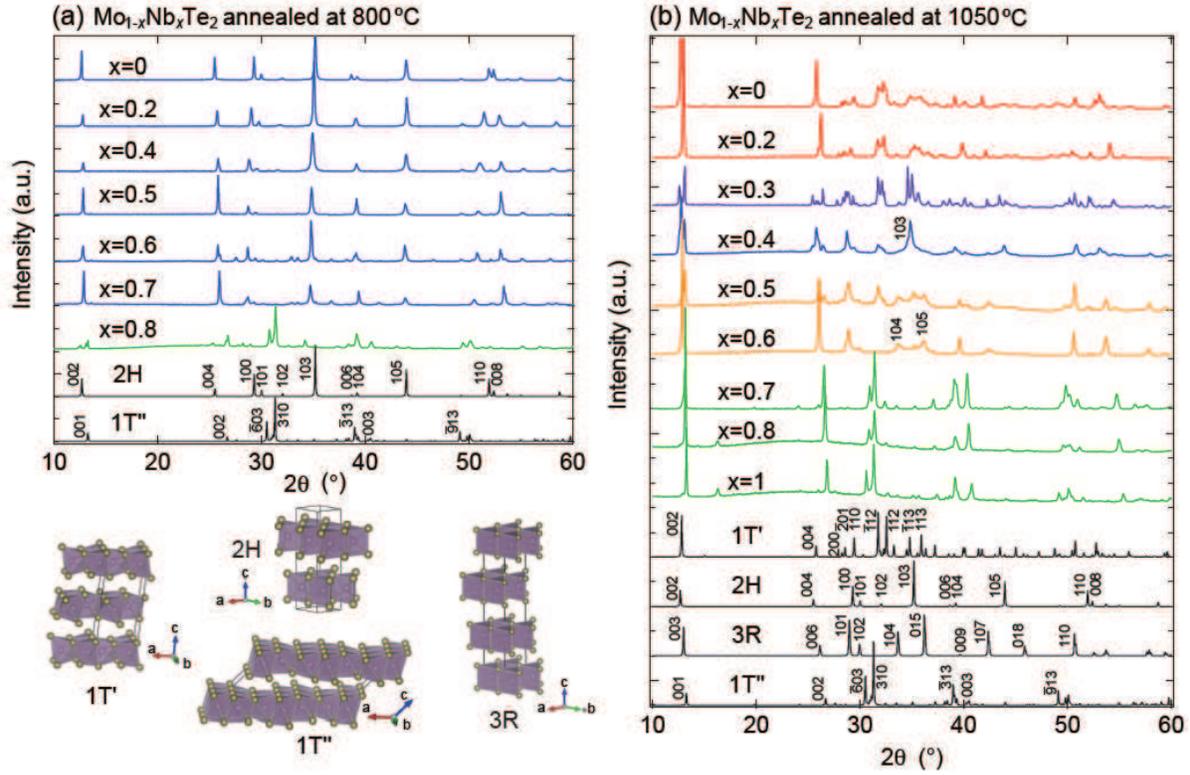}
 \caption{\label{Fig1}{(a) X-ray diffraction patterns of Mo$_{1-x}$Nb$_{x}$Te$_{2}$ obtained by annealing at 800 $^{\circ}$C, and (b) at 1050 $^{\circ}$C. Simulated x-ray diffraction patterns with reflection indexes are displayed at the bottoms. The parameters of atomic coordinations for the simulation were taken from Refs. [\cite{Brown,Towle}]. Schematic crystal structures of Mo$_{1-x}$Nb$_{x}$Te$_{2}$ are shown}. 
 }
 \end{figure}

 \begin{figure}
 \includegraphics[keepaspectratio,width=14 cm]{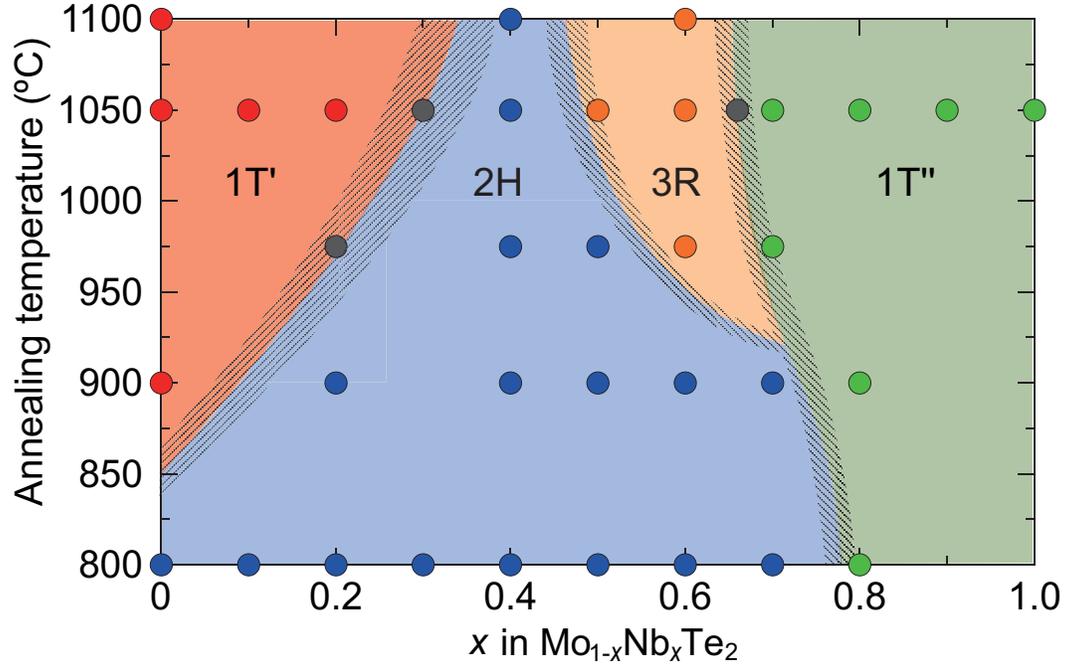}
 \caption{\label{Fig2}{Structural phase diagram of Mo$_{1-x}$Nb$_{x}$Te$_{2}$ as functions of the Nb concentration $x$ and annealing temperature. The areas, where the adjacent phases are coexisting, are shaded.}
 }
 \end{figure}
 
 \begin{figure}
 \includegraphics[keepaspectratio,width=16 cm]{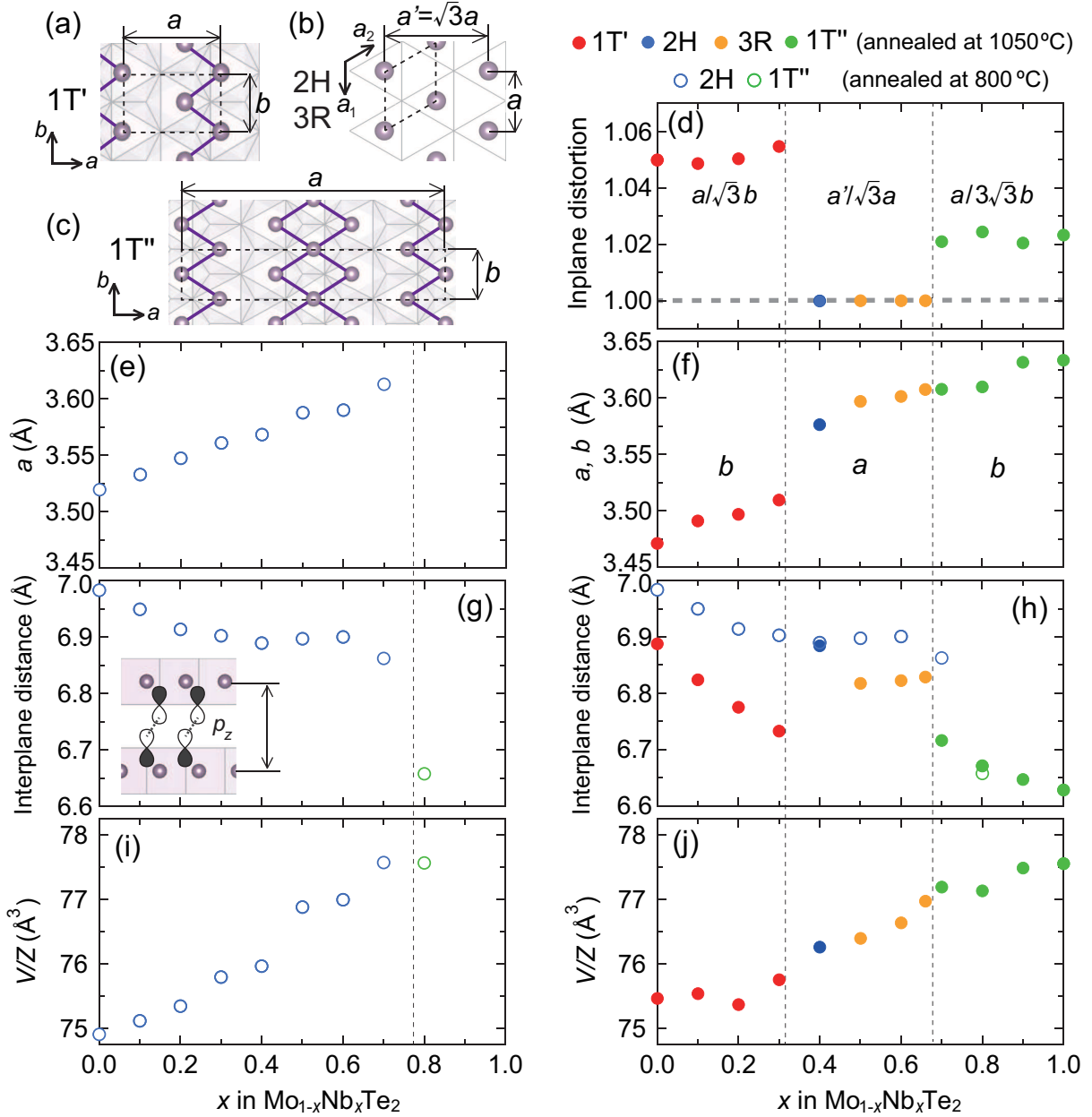}
 \caption{\label{Fig3}{Crystal structure of a single layer projected on the (001) plane for (a) 1T', (b) 2H and 3R, and (c) 1T". The chain-like arrays of Mo-Mo bonds in 1T' and 1T" polytypes are shown by thick lines. The size of the unit cell in (001) plane is indicated by broken lines. (d) Inlayer distortion defined by $a/\sqrt{3}b$ for 1T', $a'/\sqrt{3}a$=1 for 2H and 3R, and $a/3\sqrt{3}b$ for 1T", (e)(f) inplane lattice constants, (g)(h) interplane distance, and (i)(j) unit cell volume $V$ divided by number of atoms per unit cell $Z$ for Mo$_{1-x}$Nb$_{x}$Te$_{2}$ as a function of $x$}. Open circles in (e)(g)(h)(i) and closed circles in (d)(f)(h)(j) denote the data for samples annealed at 800 $^{\circ}$C and 1050 $^{\circ}$C, respectively. The definition of interplane distance and the schematic illustration of covalent bonds between Te $p_z$ orbitals are shown as an inset of (g).
 }
 \end{figure}
 
 \begin{figure}
 \includegraphics[keepaspectratio,width=16 cm]{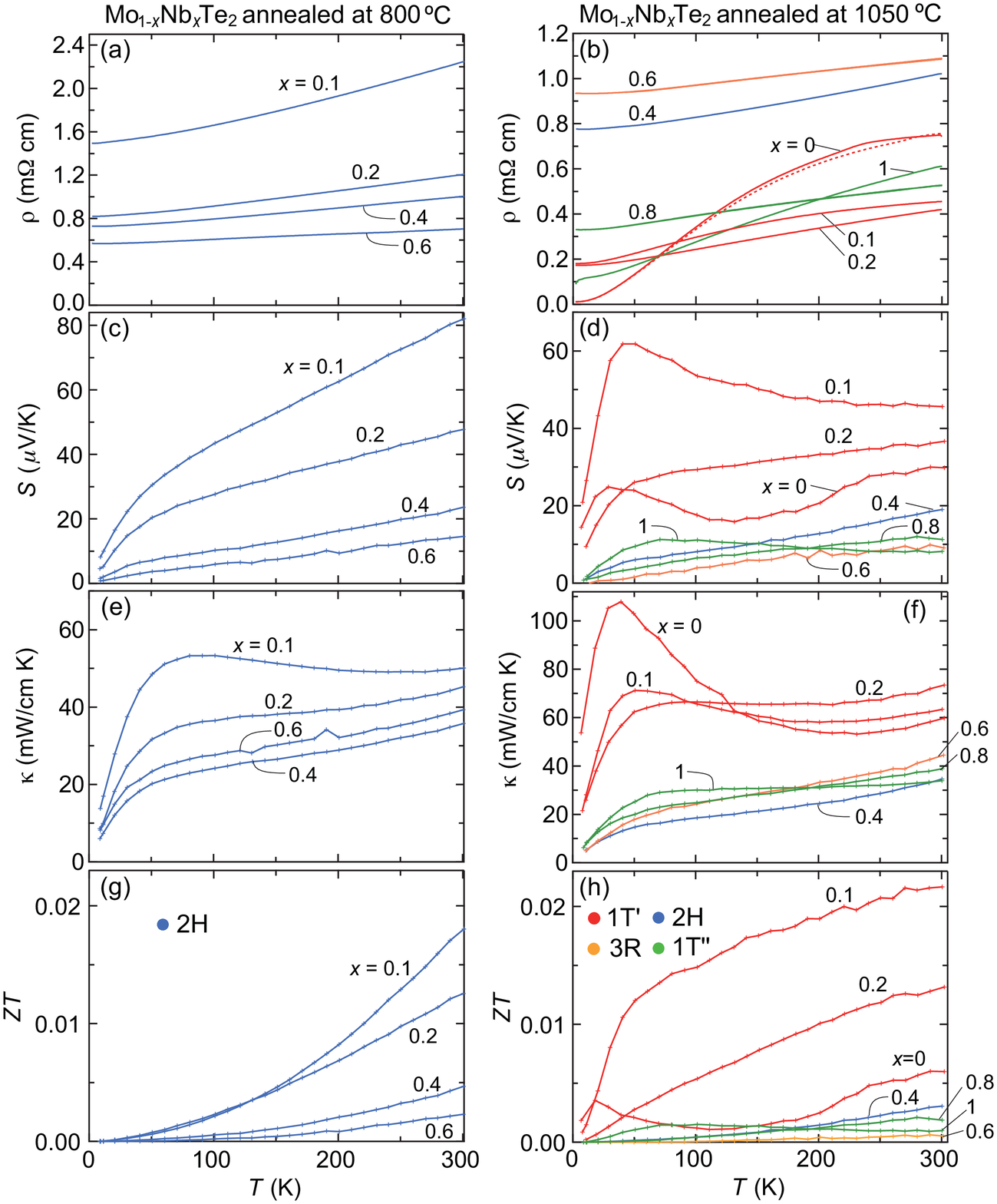}
 \caption{\label{Fig4}{Temperature dependence of (a)(b) resistivity, (c)(d) Seebeck coefficient, (e)(f) thermal conductivity, and (g)(h) figure of merit $ZT$ of Mo$_{1-x}$Nb$_{x}$Te$_{2}$ obtained by annealing at 800 $^{\circ}$C(a)(c)(e)(g) and at 1050 $^{\circ}$C (b)(d)(f)(h). All the data were measured on cooling, except for the dotted line for resistivity of 1T'-MoTe$_2$ in (b).}  
 }
 \end{figure}
 
  \begin{figure}
 \includegraphics[keepaspectratio,width=16 cm]{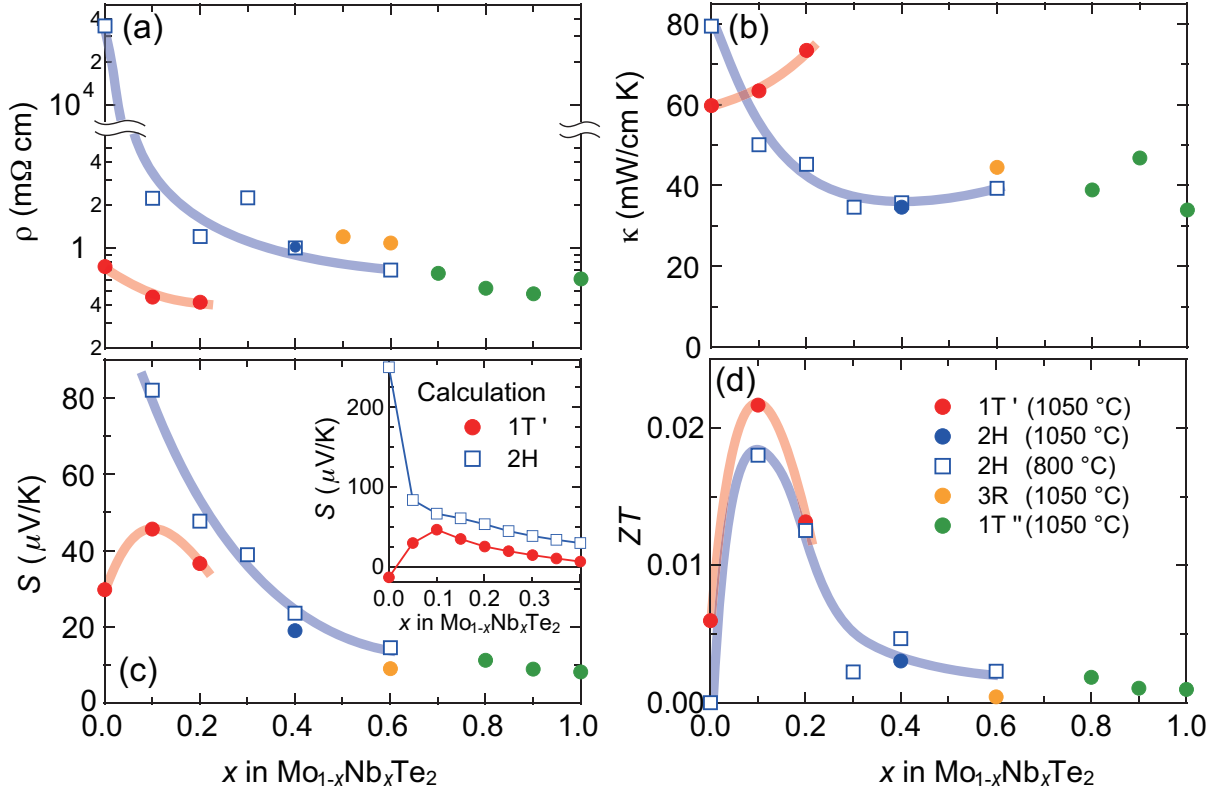}
 \caption{\label{Fig5}{Composition dependent (a) resistivity, (b) thermal conductivity, (c) Seebeck coefficient, (d) figure of merit $ZT$ of Mo$_{1-x}$Nb$_{x}$Te$_{2}$ at 300 K. Blue and red lines are guides to the eyes for 1T' and 2H polytypes, respectively. Calculated seebeck coefficients for 1T'- and 2H-Mo$_{1-x}$Nb$_{x}$Te$_{2}$ at 300 K are shown in the inset of (c).}
 }
 \end{figure}
 
 \begin{figure}
 \includegraphics[keepaspectratio,width=12 cm]{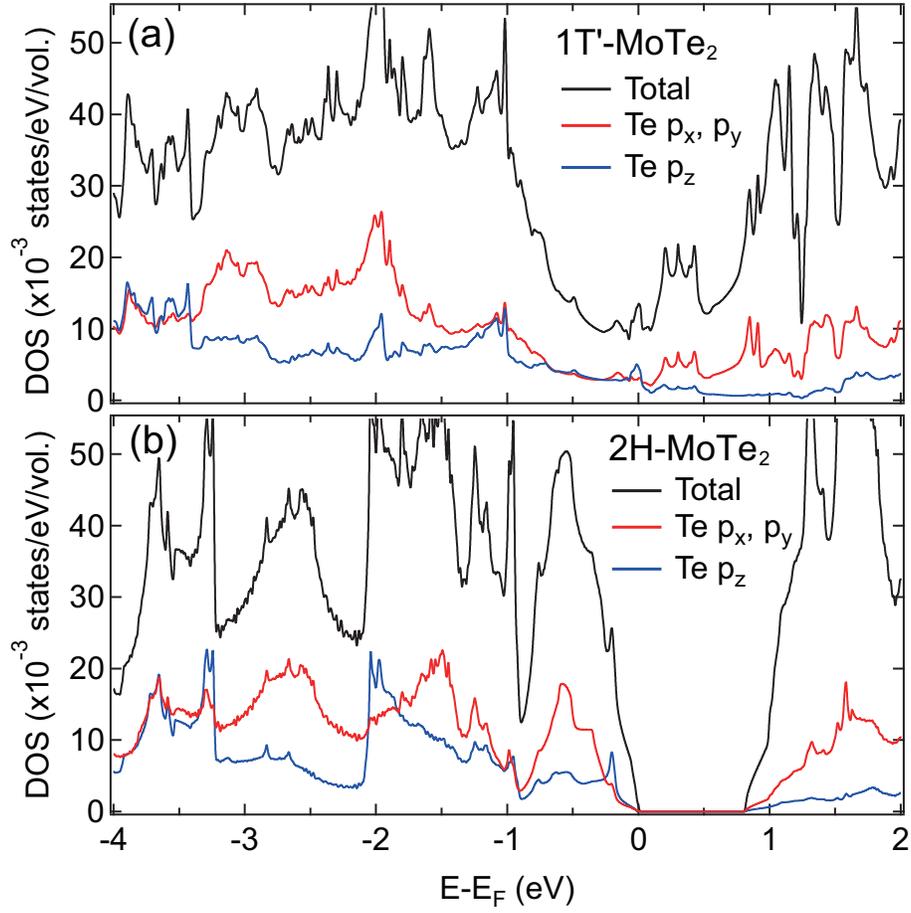}
 \caption{\label{Fig6}Total and projected DOS curves of (a) 1T'-MoTe$_{2}$ and (b) 2H-MoTe$_{2}$. 
 }
 \end{figure}
 
\end{document}